\PassOptionsToPackage{
  pdfstartview=FitH,
  breaklinks=true,
  bookmarks=true,
  colorlinks=true,
  anchorcolor=black,
  citecolor=blue,
  filecolor=black,
  menucolor=black,
  urlcolor=blue,
  linkcolor=blue
}{hyperref}
\documentclass[aps,10pt,prx,twocolumn,superscriptaddress,noeprint,longbibliography,floatfix]{revtex4-2}

\usepackage{graphicx}
\usepackage{bm}
\usepackage{amsmath, amssymb}
\usepackage{booktabs}
\usepackage{mathrsfs}
\usepackage{makecell}
\usepackage[normalem]{ulem}
\usepackage{pifont}
\usepackage{placeins}
\usepackage{lipsum}

\usepackage[mathscr]{euscript}
\usepackage{orcidlink}
\usepackage{multirow}
\usepackage{physics}
\usepackage{xcolor}
\usepackage{siunitx}
\usepackage{placeins}
\usepackage{blindtext}
\usepackage{longtable}
\usepackage{titlesec}
\usepackage{nicefrac}
\usepackage{appendix}
\usepackage{mathtools}

\newcommand{\eg}{\textit{e.g.}}

\newcommand{\HTC}{H_{\mathrm{TC}}}

\newcommand{\figref}[2]{\hyperref[#1]{\autoref*{#1}(#2)}}
\newcommand{\aref}[1]{\hyperref[#1]{App.~\ref*{#1}}}
\newcommand{\HCM}{H_{\mathrm{M}}}
\newcommand{\HBH}{H_{\mathrm{BH}}}

\newcommand{\Zp}{\mathbb{Z}_p}

\def\equationautorefname~#1\null{Eq. (#1)\null}

\newcommand{\appref}[1]{\hyperref[#1]{App.~\ref*{#1}}}
\renewcommand{\eg}[1]{\textit{e.g.}}
\usepackage[all]{hypcap} %
\usepackage{physics}
\renewcommand{\ket}[1]{\vert #1\rangle}

\renewcommand{\expval}[1]{\langle\,#1\,\rangle}

\renewcommand{\Zp}[0]{\mathbb{Z}_p}
\newcommand{\CTC}[2]{\mathbb{Z}_p ^{( #1, #2 )}}
\newcommand{\HoT}[2]{\mathrm{HoT}_p^{( #1, #2 )}}

\newcommand{\modp}[1]{#1\, \mathrm{mod} \,p}

\usepackage[backgroundcolor=green!40,textsize=tiny]{todonotes}

\begin{document}

\title{Hall-on-Toric: Descendant Laughlin state in the chiral $\mathbb{Z}_p$ toric code}


\author{Robin Sch\"afer \orcidlink{0000-0001-9728-2371}}
\affiliation{Department of Physics, Boston University, Boston, Massachusetts, 02215, USA}
\affiliation{Helmholtz-Zentrum Berlin f\"ur Materialien und Energie, Hahn-Meitner-Platz 1, 14109 Berlin, Germany}
\affiliation{Dahlem Center for Complex Quantum Systems and Fachbereich Physik, Freie Universit\"at Berlin, Arnimallee 14, 14195 Berlin, Germany}

\author{Claudio Chamon \orcidlink{0000-0002-8275-2024}}

\affiliation{Department of Physics and Astronomy, Purdue University, West Lafayette, Indiana, 47907, USA}
\affiliation{Purdue Quantum Science and Engineering Institute, Purdue University, West Lafayette, Indiana, 47907, USA}
\author{Chris R. Laumann \orcidlink{0000-0001-8979-756X}}
\affiliation{Department of Physics, Boston University, Boston, Massachusetts, 02215, USA}
\affiliation{Max-Planck-Institut f\"{u}r Physik komplexer Systeme, 01187 Dresden, Germany}

\begin{abstract} 
We demonstrate that the chiral $\mathbb{Z}_p$ toric code -- the quintessential model of topological order --- hosts additional, emergent topological phases when perturbed: descendant fractional quantum Hall-like states, which we term \textit{Hall-on-Toric}.
These hierarchical states feature fractionalized $\mathbb{Z}_p$ charges and increased topological ground-state degeneracy.
The Hall-on-Toric phases appear in the vicinity of the transitions between deconfined $\mathbb{Z}_p$ phases with different background charge per unit cell, in a fixed non-trivial flux background.
We confirm their existence through extensive infinite density matrix renormalization group (iDMRG) simulations, analyzing the topological entanglement entropy, entanglement spectra, and a generalized Hall conductance. 
Remarkably, the Hall-on-Toric states remain robust even in the absence of $U(1)$ symmetry. 
Our findings reinforce the foundational interpretation of star and plaquette defects as magnetic and electric excitations, and reveal that this perspective extends to a much deeper level.
\end{abstract}

\maketitle

\begin{figure}[t]
    \centering
    \includegraphics[width=\columnwidth]{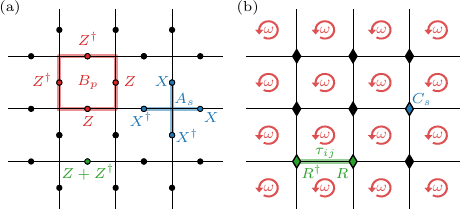}
    \includegraphics[width=\columnwidth]{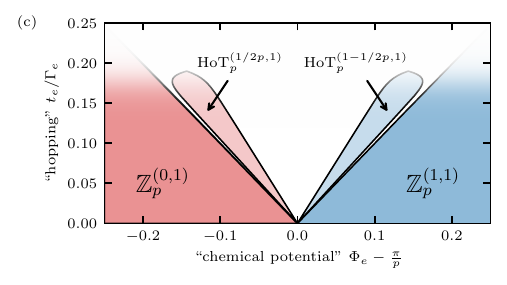}
    \caption{
        (a) The $\mathbb{Z}_p$ degrees of freedom of the chiral toric code reside on the edges of the square lattice, \autoref{eq:H0}. 
        (b) The $e$ and $m$ excitations live on the sites of the lattice and dual, respectively. 
        The $e$ dynamics can be mapped to a pure $\mathbb{Z}_p$ matter model in a fixed background flux $\expval{B_p}=\omega$, \autoref{eq:Hgauge}.
        (c) Schematic phase diagram illustrating the gapped Hall-on-Toric phases (lobes) in the vicinity of the transition between different background charge states ($\CTC{0}{1}$ and $\CTC{1}{1}$) of the $\Zp$ theory with fixed background flux. The couplings are defined in \autoref{eq:HZp} and play the role of $\Zp$ charge ``hopping'' and ``chemical potential'' in the matter model, \autoref{eq:Hgauge}. 
    }
    \label{fig:1}
\end{figure}

In what sense are the emergent electric charges in a toric code like familiar electrons? For example, in a magnetic field due to a background of emergent magnetic charges, would they undergo cyclotron motion?  If one could engineer a large ``density'' of the charges, would further interactions then produce quantum Hall physics?

Here, by tuning into a regime where the ground state hosts a finite density of electric and magnetic charges, we answer these questions in the affirmative using a surprisingly simple canonical model for topological phases --- the $\mathbb{Z}_p$ toric code illustrated in \figref{fig:1}{a}~\cite{bullock_qudit_2007,schulz_breakdown_2012,slagle_quantum_2017,vijay_isotropic_2017,watanabe_ground_2023}. 
We find a ground state phase in which the emergent electric charges form a $\nu=1/2$-type Bosonic Laughlin state~\cite{laughlin_anomalous_1983}, which we refer to as the Hall-on-Toric state. 

A major distinction between the Hall-on-Toric states and the usual fractional quantum Hall (FQH) states~\cite{laughlin_quantized_1981,tsui_two_1982,laughlin_anomalous_1983,haldane_fractional_1983,haldane_nonlinear_1983,haldane_many-particle_1985} lies in the role of charge conservation. 
In the latter, the underlying electronic charge is conserved, and the chiral edges required by the topological order carry dissipationless charge currents. Historically, those currents underlie the discovery of the FQH states in transport experiments~\cite{klitzing_new_1980,tsui_two_1982} and remain the most robust signature of the effect. In the Hall-on-Toric state, the underlying charges are only conserved modulo $p$ as they are emergent charges of a $\mathbb{Z}_p$ theory. Chiral edges remain, but they carry only thermal currents.

Just as in the hierarchical constructions of FQH states, the topological order of the Hall-on-Toric state can be understood by assuming the $\mathbb{Z}_p$ charges of the toric code form a Hall state of their own. 
In the low-energy Chern-Simons description, the $\mathbb{Z}_p$ toric code is described by the Lagrangian,
\begin{align}
    \mathcal{L}_{TC} = \frac{p}{2\pi} \,\epsilon\, a \partial b
\end{align}
where the $a$ and $b$\; $U(1)$ gauge fields couple to the $e$ and $m$ quasiparticles, respectively, and $\varepsilon$ is the Levi-Civita symbol. Placing the $e$ quasiparticles in a $\nu = 1/2$ state, one obtains
$$\mathcal{L}_{HoT} = \frac{p}{2\pi}\, \epsilon\, a \partial b + \frac{1}{2\pi}\, \epsilon\, a \partial c + \frac{2}{4\pi} \,\epsilon\, c \partial c$$
where $c$ is an additional $U(1)$ gauge field whose fluxes represent the $e$ quasiparticles. 
Consequences that follow include~\cite{wen_quantum_2007,fradkin_field_2013,simon_topological_2023}: (i) the topological degeneracy (on a torus) of $2p^2$; (ii) a chiral central charge $c = 1$, or, equivalently, thermal Hall conductance $\kappa_{xy}/T  = \frac{\pi^2 k_b^2}{3h}$; and (iii) the existence of a quasiparticle carrying $1/2$ of the toric code $\mathbb{Z}_p$ charge. 
In particular, as the underlying $e$ is only conserved modulo $p$, we have that $2p-1$ of the $e/2$ quasiparticles are equivalent to one quasihole.

\begin{figure}[t]
    \centering    
    \includegraphics[width=1\columnwidth]{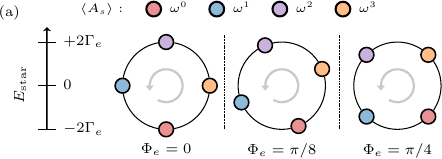}
    \includegraphics[width=1\columnwidth]{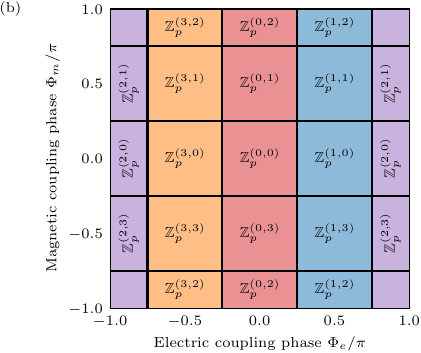}
    \caption{
        (a) The spectrum of $e^{-i\Phi_e}A_s + \text{h.c.}$ (see \autoref{eq:E_em}) can be understood as the height of the $p$ cars (colored eigenstates with eigenvalues the $p$'th roots of unity) in a Ferris wheel rotated by $\Phi_e$ (here, $p=4$).  
        At $\Phi_e=\pi/4$, a first-order phase transition in the ground state occurs, as the lowest car switches from the red to the blue. 
        (b) Phase diagram of the solvable Hamiltonian $H_0$ in \autoref{eq:H0} ($t_e = 0$). The colors (constant in columns) represent the value of $\langle A_s\rangle$ and refer to the lowest energy state in (a). 
        The value from the plaquette operator $\langle B_p\rangle=\omega^g$ is constant across rows. 
    }\label{fig:phase_diagram_unpert}
\end{figure}

These notable long-wavelength consequences only appear provided the topological state \emph{is} stabilized on the lattice. 
While the Hall-on-Toric states do not appear in the phase diagram of the exactly solvable $\mathbb{Z}_p$ Toric code model (see \figref{fig:phase_diagram_unpert}{b}), we show that they appear asymptotically close to it in the sense that even weak solvability-breaking perturbations are sufficient.
The Hall-on-Toric states intervene near the transitions between  $\mathbb{Z}_p$ deconfined phases with different background charge $q$ per unit cell, so long as there is non-trivial background flux $g$ per unit cell to break time reversal symmetry.
We denote the reference Toric-code phases by $\CTC{q}{g}$ and, following this notation, the descendant Hall-on-Toric phases by $\HoT{q}{g}$, with $q$ taking fractional values.


\paragraph*{Model.}
The $\mathbb{Z}_p$ generalization of the toric code~\cite{dennis_topological_2002,kitaev_fault-tolerant_2003} is built from $p$-dimensional clock degrees of freedom residing on the edges of a square lattice (see \figref{fig:1}{a})~\cite{bullock_qudit_2007,schulz_breakdown_2012,slagle_quantum_2017,vijay_isotropic_2017,watanabe_ground_2023}. 
The generalized Pauli operators $X$  and $Z$ are defined with respect to an orthonormal basis $\ket{s}$, $s=0,\dots,{p-1}$,
\begin{align}
    Z\ket{s} = \omega^s\ket{s}\text{ and }X\ket{s} = \ket{\modp{s-1}}
\end{align}
where $\omega=e^{2\pi i/p}$. 
The ``clock'' $Z$ and ``shift'' $X$ satisfy the generalized commutation relation: $XZ=\omega ZX$.
The plaquette and star operators are defined by,
\begin{align}
	B_p =\vcenter{\hbox{\tikz[baseline=(char.base)]{\node[shape=rectangle,draw,minimum size=20pt] (char) {};
				\draw (char.north) node[above] {$Z^\dagger$};
				\draw (char.south) node[below] {$Z$};
				\draw (10pt,1pt) node[right] {$Z$};
				\draw (-10pt,1pt) node[left] {$Z^\dagger$};
    			\node at (10pt,0) [circle, draw, fill=black, inner sep=1pt] (char) {};
    			\node at (10pt,0) [circle, draw, fill=black, inner sep=1pt] (char) {};
    			\node at (-10pt,0) [circle, draw, fill=black, inner sep=1pt] (char) {};
    			\node at (0,10pt) [circle, draw, fill=black, inner sep=1pt] (char) {};
    			\node at (0,-10pt) [circle, draw, fill=black, inner sep=1pt] (char) {};
				}}}
	\quad\quad\quad A_s =      \vcenter{\hbox{\tikz[baseline=(char.base)]{
				\draw (0,0) -- (1,0);
				\draw (0.5,-0.5) -- (0.5,0.5);
				\node at (0,0) [circle, draw, fill=black, inner sep=1pt] (char) {};
				\node at (0.5,-0.5) [circle, draw, fill=black, inner sep=1pt] (char) {};
				\node at (0.5,0.5) [circle, draw, fill=black, inner sep=1pt] (char) {};
				\node at (1,0) [circle, draw, fill=black, inner sep=1pt] (char) {};
				\node at (1,0.2) {$X$};
				\node at (0,-0.2) {$X^\dagger$};
				\node at (0.7,-0.5) {$X^\dagger$};
				\node at (0.3,0.5) {$X$};
			}}}\label{eq:plaquette_and_star}\,.
\end{align}
and, just as in the usual toric code, the full set of $A_s$ and $B_p$ operators mutually commute.

The commuting $\Zp$ toric code Hamiltonian is familiar
\begin{align}
	H_0 = -\Gamma_m e^{-i\Phi_m}\sum_p  B_p - \Gamma_e e^{-i\Phi_e} \sum_s A_s +\text{h.c.} \,.\label{eq:H0}
\end{align}
but for the key feature that $A_s$ and $B_p$ are unitary but not Hermitian for $p>2$. 
Thus, the couplings can be complex, which we write here in polar form with positive amplitudes, $\Gamma_m$ and $\Gamma_e$, and phases $\Phi_m$ and $\Phi_e$.
We note that $\Phi_m$ is odd under parity and $\Phi_e$ is odd under time-reversal (see \aref{app:model}) and thus $H_0$ defines the \emph{chiral} $\Zp$ toric code.
The sums $\sum_p$ and $\sum_s$ run over all plaquettes and stars, respectively. 

As a function of the coupling phases, the Hamiltonian $H_0$ exhibits $p^2$ distinct $\mathbb{Z}_p$ deconfined phases, $\CTC{q}{g}$ (see \figref{fig:phase_diagram_unpert}{b}). 
Each phase has a topological degeneracy of $p^2$ on the torus, a topological entanglement entropy of $-\log p$, and host $e$ and $m$ anyons with mutual statistics~\cite{wilson_confiement_1974,bullock_qudit_2007,watanabe_ground_2023}.
They are distinguished by the background charge $q$ and flux $g$ per unit cell; for the commuting model, these equivalently give the expectation values of the star and plaquette operators in the ground state: $\langle A_s \rangle = \omega^{q}$ and $\langle B_p \rangle = \omega^{g}$. 
For example, the ground state value of $\langle A_s\rangle$ is selected by finding the integer value of $q$ that minimizes
\begin{align}
    E_{\mathrm{star}}(q) = -2\Gamma_{e}\cos\left(\frac{2\pi q}{p}-\Phi_e\right)
    \label{eq:E_em}
\end{align}
for given $\Phi_e$.
We illustrate this relation by a rotating clock in \figref{fig:phase_diagram_unpert}{a} showing the energy $E_{\mathrm{star}}$ of the $p=4$ eigenvalues of $A_s$, which are equidistantly placed on the unit circle. 
At $\Phi_e=\pi/4$, the system undergoes a first-order transition, where $\expval{A_s}$ goes from $\omega^0$ (red) to $\omega^1$ (blue): $\CTC{0}{g}\rightarrow\CTC{1}{g}$. 
An analogous discussion holds for the plaquette operator $B_p$ and the parameter $\Phi_m$.

We perturb the solvable model with a uniform field coupling to $Z$,
\begin{align}
    \HTC = H_0 - t_e\sum_e \left(Z_e + Z_e^\dagger\right)
    \label{eq:HZp}\,.
\end{align}
The sum runs over all edges $e$ of the square lattice, and we take the parameter $t_e > 0$ to be real.
The full model is illustrated in \figref{fig:1}{a}.

Since the perturbation commutes with the plaquette operators $B_p$, the model remains diagonal in plaquette flux, even as the charges begin to fluctuate.
To focus on these dynamics, we extend the  Hilbert space by introducing $\Zp$ matter degrees of freedom, $C_s$, on the sites of the square lattice, subject to the local Gauss law constraint that $C_s = A_s$ (see~\aref{app:model} for more details). 
In the extended gauge-matter representation, the gauge degrees of freedom are static and we obtain a pure $\Zp$ matter model,
\begin{align}
    \HCM &= -\Gamma_e\sum_s e^{-i\Phi_e}{C}_s- t_e\sum_{\langle i,j\rangle} \tau_{ij}\; {R}^\dagger_{i}{R}_{j}+\text{h.c.}\,.\label{eq:Hgauge}
\end{align}
Here, $R_s$ is $\Zp$ conjugate to $C_s$ and $\tau_{ij}$ is a fixed gauge background determined by the $B_p$ flux sector of interest.
The matter model is a $\mathbb{Z}_p$ version of the Harper-Hofstadter Hamiltonian~\cite{harper_single_1955,hofstadter_energy_1976}, see \figref{fig:1}{b}.
We emphasize that Eq.~\eqref{eq:Hgauge} is an exact rewriting of Eq.~\eqref{eq:HZp} available so long as the plaquette flux is static.

Consider the transition between two toric code phases $\CTC{0}{1}$ with no background charge per unit cell and $\CTC{1}{1}$ with one background charge per unit cell.
At $t_e=0$, the transition is direct, taking place at $\Phi_e=\pi/p$ where the energy of the two charge states is degenerate as illustrated by the ``ferris wheel'' in \figref{fig:phase_diagram_unpert}{a}. 
For $t_e \ll \Gamma_e$, the other charge states are energetically inaccessible and may be projected out. 
The resulting effective Hamiltonian is a Hofstadter model with hardcore Bosons,
\begin{align}
    \HBH &= -\mu \sum_s n_s - t_e \sum_{\langle i,j\rangle} \tau_{ij} a^\dagger_ia_{j}+\text{h.c.}\label{eq:H_BH}\,.
\end{align}
Here, $a^\dagger_s$ and $a_s$ are the Bosonic creation and annihilation operators on site $s$ and we have the charge state $C_s = \omega^{n_s}$. 
The effective chemical potential $\mu =-4 \Gamma_e\sin\left(\frac{\pi}{p}\right)\sin\left(\frac{\pi}{p}-\Phi_e\right)$ vanishes at the transition linearly with $\Phi_e$.

In contrast to the unprojected $\Zp$ model, which only conserves the charge modulo $p$, the effective Bosonic model at small $t_e$ exhibits an emergent $U(1)$ symmetry. 
Within a Schrieffer-Wolff treatment, virtual processes that eventually break the $U(1)$ are suppressed with $\left({t_e}/{\Gamma_e}\right)^{p-1}$ for $p \leq 5$. 
For $p \geq 6$, the suppression becomes even stronger, since additional virtual processes are needed when $p-1$ exceeds the coordination number of the lattice.

In analogy to the Mott lobes~\cite{fisher_boson_1989} in the Bose-Hubbard problem, we can estimate the splitting of the transition by comparing the single particle energy~\cite{harper_single_1955} $\Delta_p$ (which is proportional to $t_e$) with the chemical potential~\footnote{For $p=4$, the ground-state energy is $\Delta_4 = -2\sqrt{2}$, and we find $\kappa_4 = \arcsin\left(t_e/\Gamma_e\right)$.}. This gives us the scale, $\kappa_p$, on which the toric code phases, $\CTC{0}{1}$ and $\CTC{1}{1}$, give way to intermediate filling,
\begin{align}
    \kappa_p = \arcsin\left(\frac{-\Delta_p}{4\Gamma_e\sin\left({\pi}/{p}\right)}\right)\approx \frac{-\Delta_p}{4\Gamma_e\sin\left({\pi}/{p}\right)}\label{eq:kappa_p}\,.
\end{align}

\begin{figure}[t]
    \includegraphics[width=\columnwidth]{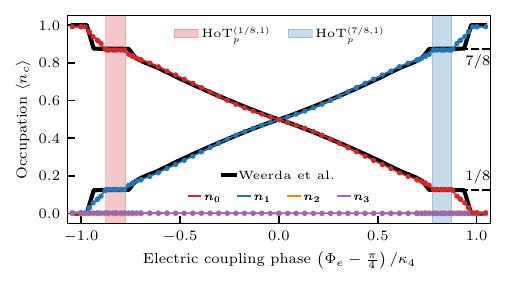}
    \caption{Charge occupation computed with iDMRG (bond dimension $\chi=512$) for $p=4$ and $L_y=8$, $L_x=4$ (iDMRG unit cell), and $t_e/\Gamma_e=0.1$ across the transition at $\Phi_e = \pi/4$ in units of $\kappa_p$ from \autoref{eq:kappa_p}. 
    The colors represent the occupation $n_c$ of the individual charge states \autoref{eq:nc} within the matter model in \autoref{eq:Hgauge} and match \autoref{fig:phase_diagram_unpert}. 
    The data in black refers to a number-conserving hardcore Bosonic Hofstadter model from Refs.~\cite{weerda_data_2023,weerda_fractional_2024} (iPEPS). 
    The plateaus correspond to the phases $\HoT{1/8}{1}$ and $\HoT{7/8}{1}$.}
    \label{fig:charge_occupation}
\end{figure}

The Bosonic Hofstadter model has been studied using different numerical methods and found to host a variety of fractional quantum Hall states. 
It has been explored using exact diagonalization~\cite{bai_bosonic_2018,umucalilar_fractional_2010,soerensen_fractional_2005,moeller_composite_2009,hafezi_fractional_2007,hafezi_characterization_2007,sterdyniak_particle_2012,geraedts_emergent_2017,zeng_bosonic_2016}, DMRG on finite lattices~\cite{rosson_bosonic_2019,palm_bosonic_2021}, iDMRG on cylindrical geometries~\cite{he_realizing_2017,boesl_characterizing_2022,geraedts_emergent_2017}, tree-tensor networks~\cite{gerster_fractional_2017}, (i)PEPS~\cite{macaluso_charge_2020,weerda_fractional_2024}, and various variational and mean-field approaches~\cite{bai_bosonic_2018,umucalilar_fractional_2010,huegel_anisotropic_2017,natu_competing_2016,ledinauskas_universal_2025}.
While these studies suggest the existence of certain Bosonic Hall states --- predominantly the Bosonic Laughlin state~\cite{laughlin_anomalous_1983} at filling $\nu = 1/2$ --- the stability of more exotic fractional states in the thermodynamic limit remains uncertain. In particular, topologically trivial density waves and other crystalline states at specific fillings compete with the proposed topologically ordered Hall states. For example, the competition with the Tao-Thouless state~\cite{tao_fractional_1983} has been highlighted in several works~\cite{rezayi_laughlin_1994,bergholtz_pfaffian_2006,seidel_abelian_2006}. This competition is especially relevant in cylindrical geometries, where the circumference becomes comparable to the magnetic unit cell length.
Thus, the stability of other filling fractions in the thermodynamic limit remains an open question, as results depend on both the method and system size. Beyond the $\nu=1/2$ state, some studies~\cite{he_realizing_2017,weerda_fractional_2024,boesl_characterizing_2022} suggest the possibility of a Bosonic integer quantum Hall state at $\nu=2$, which is constructed from composite fermions~\cite{jain_composite_1989,moeller_composite_2009}. 
Given the increased complexity of simulating the $\mathbb{Z}_p$ matter model, we concentrate on the Laughlin state at half filling.

\paragraph*{Results.} 
We numerically study the matter model in \autoref{eq:Hgauge} with a fixed background flux of $\langle B_p\rangle=\omega$ by placing the system on an infinite cylinder, and computing its ground state using iDMRG~\cite{hauschild_efficient_2018}. Directly simulating the full toric-code Hamiltonian in \autoref{eq:HZp} is numerically much more demanding, as it involves twice the number of sites and includes four-body operators.
We evaluate the following quantities: (\emph{i}) $\Zp$ charge occupation, (\emph{ii}) topological entanglement entropy, (\emph{iii}) generalized ``Hall conductance'', and (\emph{iv}) the entanglement spectrum.

\paragraph*{(i) $\Zp$ charge occupation:} The perturbation $t_e$ splits the first-order transition between the $\CTC{0}{1}$ and $\CTC{1}{1}$ phases and gives rise to a variety of competing phases in between. 
\autoref{fig:charge_occupation} shows the normalized $\Zp$ charge occupation for $p = 4$ as a function of $\Phi_e$ which refers to a horizontal line cut in \figref{fig:1}{c} at $t_e/\Gamma_e=0.1$:
\begin{align}
    n_c = \frac{1}{L_x L_y} \sum_s P_{sc}\,,\label{eq:nc}
\end{align}
where $P_{sc}$ is the projector onto the state $\ket{c}$ at site $s$. 
The charge occupation $n_c$ is related to the star operator: 
${A_s}=\omega^ {q_s}${ with } $q_s=\sum_{{c}}cP_{sc}$. For comparison, we show data from Refs.~\cite{weerda_data_2023,weerda_fractional_2024}, which have simulated the (hardcore) Bosonic Hofstadter model using iPEPS for $p=4$. Additional data for different $p$, $t_e/\Gamma_e$, and scaling with $L_y$ is shown in \aref{sec:data}.

\begin{figure}[t]
    \centering
    \includegraphics[width=\columnwidth]{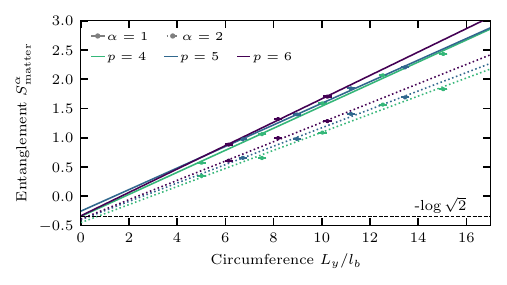}
    \caption{Topological entanglement entropy of the matter model (\autoref{eq:Hgauge} with $t_e/\Gamma_e=0.05$) ground state measured for different circumferences $L_y$ in units of the magnetic length $l_b={p}/{2\pi}$ along the cut in a cylinder using iDMRG ($\chi\geq 1024$, and $L_x=p$). The black, blue, and red data points correspond to $p=4$, $p=5$, and $p=6$, calculated at $\Phi_e = 0.2368\pi$, $\Phi_e=0.1825\pi$, and $\Phi_e = 0.1891\pi$, respectively.
    $S_0$ and $S_1$ refer to the von Neumann and first Rényi entropy. 
    The expected topological entanglement entropy of the Hall-on-Toric state is $-\log \sqrt{2}$ (from the Laughlin state) plus the gauge sector contribution $-\log p$.}
    \label{fig:entanglement_entropy}
\end{figure}

For $\Phi_e\lesssim\frac{\pi}{p}-\kappa_p$, the system is in the $\CTC{0}{1}$ phase, which appears as a trivial paramagnet within the matter model composed solely of red charges with $n_0 = 1$ and zero otherwise. At $\Phi_e\approx\frac {\pi}{p} - \frac{5}{6}\kappa_p$, a plateau emerges at fractional charge occupation. 
The blue charges have increased to a finite density, $n_1=1/(2p)$, and form the $\nu=1/2$ state, embedded in a background of red charges at density, $n_0=(2p - 1)/(2p)$~\cite{huber_topological_2011}. 
We refer to this state as $\HoT{1/2p}{1}$. An analogous Hall state, $\HoT{1-1/2p}{1}$, is formed at $\Phi_e\approx\frac {\pi}{p} + \frac{5}{6}\kappa_p$, where the roles of the two charge states are reversed.
The model exhibits charge-anti-charge symmetry around the transition ($\Phi_e\longleftrightarrow 2\pi/p-\Phi_e$), which is composed of charge conjugation and a global charge shift, as defined in \aref{app:model}. This refers to the particle-hole symmetry in the Hofstadter model.

\begin{figure}[t]
    \includegraphics[width=\columnwidth]{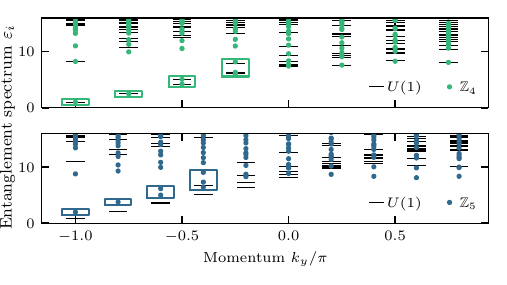}
    \caption{Entanglement spectrum computed in the matter model (\autoref{eq:Hgauge} with $t_e/\Gamma_e=0.05$) with iDMRG ($\chi=1024$, $L_x=p$, and $L_y=2p$) for $p=4$ ($\Phi_e=0.2368\pi$) and $p=5$ ($\Phi_e=0.1825\pi$) in (a) and (b), respectively. The plots display the entanglement energies $\varepsilon_i = -\log \Lambda_i$ in the $c=0$ sector, where $\Lambda_i$ are the Schmidt values obtained from a bipartition along the cylinder. The black stripes indicate the corresponding calculation in the number-conserving Hofstadter model from \autoref{eq:H_BH}. The boxes highlight the characteristic counting sequence expected for the Laughlin state.\label{fig:entanglement_spectrum}}
\end{figure}
Next, we analyze intrinsic properties of the $\HoT{1/2p}{1}$ state accessible via its Schmidt decomposition. Thereby, we use the $\mathbb{Z}_p$ symmetry (particle number conservation modulo $p$) and the translational invariance to group the Schmidt coefficients. $k_y$ refers to the $y$-momentum around the cylinder, and $c = 0, \ldots, p-1$ labels the charge differences across the cut~\cite{singh_tensor_2010,singh_tensor_2011,hauschild_efficient_2018}:
\begin{align}
     \ket{\Psi} = \sum_{k_y}\sum_{c=0}^{p-1}\sum_{i=1}^{\chi_{(k_y,c)}} \Lambda_{(k_y,c,i)} \ket{\psi^L_{(k_y,c,i)}} \otimes \ket{\psi_{(k_y,c,i)}^R}\,.\label{eq:Psi}
\end{align}

\paragraph*{(ii) Topological entanglement entropy:}The first intrinsic property we analyze is the topological entanglement entropy~\cite{levin_detecting_2006,kitaev_topological_2006}. It encodes information about Abelian excitations by extracting a universal correction, $\gamma$, to the area law. We compute the entanglement across the cut and obtain $\gamma$ from the following scaling relation in the matter model:
\begin{align} 
S_{\mathrm{matter}}^\alpha(L_y) = c_\alpha L_y + \gamma + \mathcal{O}\left(\frac{1}{L_y}\right)\,.
\end{align}
Here, $L_y$ refers to the cylinder's circumference, and $\alpha$ denotes different R\'enyi entropies. \autoref{fig:entanglement_entropy} shows the scaling of the von Neumann entropy ($\alpha=1$) and the second R\'enyi entropy ($\alpha=2$) with $L_y$ for different values of $p$. It shows a good agreement with the expected value of $\gamma=-\log\sqrt{2}$ for $\nu=1/2$, which remains unchanged across different R\'enyi entropies~\cite{dong_topological_2008,zhang_quasiparticle_2012,zaletel_topological_2013,zaletel_flux_2014}. We remark that the simulated matter model has frozen out the flux degrees of freedom by employing a fixed gauge. The entanglement entropy in the full $\mathbb{Z}_p$ model includes an additional contribution of $-\log p$ from the gauge sector.

\paragraph*{(iii) Generalized ``Hall conductance'':}We further analyze the Hall response by adiabatically threading flux ($\Theta_y$) through the cylinder~\cite{laughlin_quantized_1981}. This measure is standard for number-conserving systems within iDMRG~\cite{grushin_characterization_2015,he_obtaining_2014,zaletel_flux_2014,boesl_characterizing_2022,huber_topological_2011,he_bosonic_2015} but we generalized it to the $\Zp$ case. Using the decomposition of the ground state in \autoref{eq:Psi}, we define the weight per charge channel:
\begin{align}
    W_c(\Theta_y):= \sum_{k_y}\sum_{i=1}^{\chi_{(k_y,c)}}  \Lambda_{(k_y,c,i)}^2(\Theta_y)\,.
\end{align}
When the system is adiabatically evolved from $\Theta_y = 0$ to $\Theta_y = 4\pi$, we find that the weights shift cyclically
\begin{align}
    W_c(0) = W_{\modp{c+1}}(4\pi)\,.
\end{align}
In the conventional $U(1)$ case, a similar shift occurs, but without the cyclic constraint. Data and a comparison between the $U(1)$ and $\mathbb{Z}_p$ cases are provided in \aref{app:Hall}.

\paragraph*{(iv) Entanglement spectrum:}The last characteristic we analyze is the entanglement spectrum~\cite{li_entanglement_2008}. For the Laughlin wavefunction, an edge conformal field theory describing a $U(1)$ chiral Boson predicts a characteristic counting sequence $\{1,1,2,3, \dots\}$ in the entanglement spectrum, given by $\varepsilon_i =-2 \log \Lambda_{(k_y,c,i)} $, when plotted against $k_y$ for each $c$~\cite{grushin_characterization_2015,cincio_characterizing_2013,boesl_characterizing_2022}.
\autoref{fig:entanglement_spectrum} presents the entanglement spectrum of the matter model~\autoref{eq:Hgauge} for $p=4$ and $p=5$, comparing it with a number-conserving calculation of \autoref{eq:H_BH}. While we find excellent agreement for $p = 4$, the $U(1)$ case resolves the sequence more clearly for $p=5$. We attribute this to the increased computational complexity and anticipate further improvements at higher bond dimensions.

\paragraph*{Potential realizations.}
Recently, $\mathbb{Z}_3$ toric code order has been demonstrated on a trapped-ion quantum processor with 24 effective qutrits~\cite{iqbal_qutrit_2025}. Per-qutrit fidelities exceeded 95\%, and several braiding operations of $\mathbb{Z}_3$ charges were realized. A number of proposals~\cite{giudice_trimer_2022} have explored the realization of $\mathbb{Z}_3$-type models across diverse quantum simulation platforms, both predating the trapped-ion experiment and motivated by subsequent developments~\cite{burshtein_robus_2025}. While none of these works explicitly target the parameter regime associated with the Hall-on-Toric state, they indicate rapid progress towards a potential experimental realization.

\paragraph*{Conclusion.}
We have demonstrated that the chiral toric code in a small field hosts hierarchical topological states in which the usual deconfined charges form a Laughlin-like Hall state of their own. 
We refer to these as Hall-on-Toric states. 
Our understanding is grounded in a perturbative mapping between the $\mathbb{Z}_p$ model and the --- number-conserving --- Bosonic Hofstadter problem. 
We confirm that the Hall-on-Toric states remain stable even without particle conservation.

Our results are quite surprising, as the Hall-on-Toric state arises in such a small perturbation of a well-known exactly solvable model. 
On the other hand, this means that the Hall-on-Toric states we find are within sight of well-known bosonic Laughlin states, in which charge non-conservation is only a weak correction to spectral properties.
It would be very interesting to find ``strong'' Hall-on-Toric states which have no analogues in the conventional Hall hierarchies.
Such states would presumably take advantage of the modular nature of number conservation in a fundamental way, similarly to the parafermionic states discovered in the context of superconducting proximity effects~\cite{read_beyond_1999,xia_electron_2004,nayak_non_2008}.
We leave the search for other Hall-on-Toric states, either weak or strong, to future investigations.

The data used to generate the plots is available on Zenodo~\cite{schaefer_data_2025}.
\begin{acknowledgments}
R.S. thanks Julian Boesl for insightful discussions. C.R.L. acknowledges the Gutzwiller Fellowship at Max Planck Institute for the Physics of Complex Systems and the partial support of SFB 1143 and ct.qmat. The work of C.C. is supported by the DOE Grant DE-SC0026189. R.S. was supported by the AFOSR
(Grant no. FA9550-24-1-0121). Simulations were conducted using the TeNPy package~\cite{hauschild_efficient_2018}.
\end{acknowledgments}

\bibliography{references}
\FloatBarrier

\appendix

\section{Model}
\label{app:model}

This appendix provides additional details on the $\Zp$ toric code model (\autoref{eq:HZp} from the main text) and the matter model (\autoref{eq:Hgauge} from the main text). 
We review the derivation of the matter representation from the toric code model, and the symmetries of the $\mathbb{Z}_p$ model.
We work with periodic boundary conditions, and set at least one length ($L_x$ or $L_y$) to be divisible by $p$. 

\subsection{Toric code and gauge-matter representation}

We recapitulate the $\Zp$ toric code in physical edge variables (see \autoref{eq:HZp}),
\begin{align}
\label{eq:tp_tc_appendix}
\HTC &= H_0 + H_1\\\
H_0 &= - \Gamma_e e^{-i \Phi_e}\sum_s A_s - \Gamma_m e^{-i \Phi_m}\sum_p B_p  + \text{h.c.}\\
H_1 &= - t_e \sum_e \left(Z_e + Z_e^\dagger \right)
\end{align}
where the star and plaquette operators are
\begin{align}
	B_p &= \prod_{e \in \partial p} Z_e = \vcenter{\hbox{\tikz[baseline=(char.base)]{\node[shape=rectangle,draw,minimum size=20pt] (char) {};
				\draw (char.north) node[above] {$Z^\dagger$};
				\draw (char.south) node[below] {$Z$};
				\draw (10pt,1pt) node[right] {$Z$};
				\draw (-10pt,1pt) node[left] {$Z^\dagger$};
    			\node at (10pt,0) [circle, draw, fill=black, inner sep=1pt] (char) {};
    			\node at (10pt,0) [circle, draw, fill=black, inner sep=1pt] (char) {};
    			\node at (-10pt,0) [circle, draw, fill=black, inner sep=1pt] (char) {};
    			\node at (0,10pt) [circle, draw, fill=black, inner sep=1pt] (char) {};
    			\node at (0,-10pt) [circle, draw, fill=black, inner sep=1pt] (char) {};
				}}}
	\\ 
    A_s &=  \prod_{e \in \delta s} X_e =     \vcenter{\hbox{\tikz[baseline=(char.base)]{
				\draw (0,0) -- (1,0);
				\draw (0.5,-0.5) -- (0.5,0.5);
				\node at (0,0) [circle, draw, fill=black, inner sep=1pt] (char) {};
				\node at (0.5,-0.5) [circle, draw, fill=black, inner sep=1pt] (char) {};
				\node at (0.5,0.5) [circle, draw, fill=black, inner sep=1pt] (char) {};
				\node at (1,0) [circle, draw, fill=black, inner sep=1pt] (char) {};
				\node at (1,0.2) {$X$};
				\node at (0,-0.2) {$X^\dagger$};
				\node at (0.7,-0.5) {$X^\dagger$};
				\node at (0.3,0.5) {$X$};
			}}}\label{eq:plaquette_and_star_appendix}\,.
    \end{align}
The generalized $\mathbb{Z}_p$ Pauli operators satisfy the clock algebra,
\begin{align}X Z = \omega Z X = e^{2\pi i/ p} Z X\end{align}
and the physical Hilbert space is $\mathcal{H} = (\mathbb{C}^p)^{\otimes N_e}$ with $N_e$ the number of edges.
Here, the boundary $\partial$ and coboundary $\delta$ operators are oriented and we define $X_{-e} = X^\dagger_{e}$ and similarly for $Z$. 
The orientation only matters for $p>2$.
We note that $X_e, Z_e, A_s, B_p$ are all unitary but, for $p>2$, not Hermitian (rather they are $p$'th roots of the identity).

For $t_e=0$, we have that both $[A_s,\HTC]=0$ and $[B_p, \HTC]=0$ for all $s,p$. 
For $t_e > 0$, the plaquette operators still commute with $\HTC$, but not the star operators. 
Assuming that $\Gamma_m$ is large enough, the ground state should live in the $B_p$ sector with $B_p = \omega^g$ for some $g$ which minimizes the $B_p$ terms. 
Furthermore, the Wilson loops $W_Z[\ell] = \prod_{e \in \ell} Z_e$ for non-trivial cycles $\ell$ are also conserved and take values among the powers of $\omega$. 

To make further progress, we extend the Hilbert space to have separate gauge and matter fields by introducing additional $\mathbb{Z}_p$ clock degrees of freedom $C_s$ on the sites of the lattice with conjugate ``rotating'' operators $R_s$:
\begin{align}
    R C = \omega C R\,.
\end{align}
We impose the constraint that $C_s = A_s$, or, equivalently, that $G_s = C_s^\dagger A_s \equiv 1$ to recover the physical Hilbert space. 
In the extended system, we can rewrite the Hamiltonian
\begin{align}
H_0 &= - \Gamma_e e^{-i \Phi_e} \sum_s C_s - \Gamma_m e^{-i\Phi_m} \sum_p B_p  + \text{h.c.}\\
H_1 &= - t_e \sum_e R^\dagger_{\partial e_0} Z_e R^{\,}_{\partial e_1} + \text{h.c.}
\end{align}
where $\partial e_0$ and $\partial e_1$ are the sites at the beginning and end of edge $e$. 
Recalling the system is on a torus, the global gauge constraint directly constrains the matter variables:
\begin{align}
\prod_s G_s = \prod_s C_s^\dagger \prod_{e \in \delta s} X_e = \prod_s C_s^\dagger \equiv 1\,.
\end{align}
In the second equality, we have observed that every edge appears twice, with opposite orientation in the product, and thus the $X_e$ operators cancel. To wit, the matter degrees of freedom must be global $\mathbb{Z}_p$ charge neutral. 

This extended representation helps to solve the perturbed model as now $[Z_e, \HTC] = 0$. 
We may work in any fixed $Z_e = \tau_e$ sector we like, so long as we apply the gauge projection to obtain physical states. 
The $B_p$  eigenvalues of a flux sector of interest are given by
\begin{align}
    B_p &= \prod_{e \in \partial p} \tau_e
\end{align}
where again $\tau_{-e} = \bar{\tau}_e$. 
In our numerical computations, we typically work in Landau gauge for a uniform flux background ($g$ fluxes per unit cell),
\begin{align}
    \tau_{\langle ij\rangle} &= \begin{cases} 
    1 & \langle ij \rangle\textrm{~ horizontal} \\
    \omega^{g i_x} & \langle ij \rangle \textrm{~ vertical}
    \end{cases}\label{eq:tau_def}
\end{align}
where $i_x$ is the $x$ coordinates of site $i$.

In a fixed gauge sector, we are left with the matter Hamiltonian (cf. \autoref{eq:Hgauge})
\begin{align}
\HCM = - \Gamma_e e^{-i \Phi_e} \sum_s C_s - t_e  \sum_e \tau_e R^\dagger_{\partial e_0} R_{\partial e_1} + \text{h.c.}
\end{align}
which has global $\mathbb{Z}_p$ symmetry:
\begin{align}
    [\prod_s C_s, \HCM] = 0
\end{align}
and which we must solve in the charge neutral sector $\prod_s C_s = 1$.

With periodic boundary conditions, there are in fact $p^2$ globally distinct gauge sectors of the model with the same background $B_p$ flux patterns.
These are distinguished by the non-trivial Wilson loop operators~\cite{wilson_confiement_1974},
\begin{align}
    W_{x/y} &= \prod_{e \in \ell_{x/y}} Z_e = \prod_{e \in \ell_{x/y}} \tau_{e}\,\label{eq:wilson_tau}
\end{align}
where $\ell_x$ and $\ell_y$ are fixed paths wrapping the $x-$ and $y-$ directions, respectively.
The eigenvalues of the loop operators are $p$'th roots of unity, which can be used to label the sectors.
In the $\Zp$-deconfined phase, the $p^2$ ground states across the sectors correspond to the topological degeneracy of the phase. 
Each sector has an additional 2-fold topological ground state degeneracy in the Hall-on-Toric state.

\subsection{Global symmetries}

\paragraph{Space group---}
In terms of the physical toric code variables, we define the unitary action of the square lattice space group $G$ in the usual way for an edge field. 
To wit, any spatial transformation $g \in G$ which maps sites $i \to g(i)$, maps edges $e = \langle ij \rangle \to g(e) = \langle g(i) g(j) \rangle$.
We then define 
\begin{align}
    g : \begin{cases} 
        X_{e} \to X_{g^{-1}(e)} \\
        Z_{e} \to Z_{g^{-1}(e)} \\
    \end{cases}\,.
\end{align}
Recall that edges $e = \langle ij \rangle = -\langle j i \rangle$ have an orientation and that $X_{e} = X^{-1}_{-e} = X^{\dagger}_{-e}$ (and similarly for $Z_e$). 

Under these transformations, $A_s$ transforms like a scalar:
\begin{align}
g: A_s &\to A_{g^{-1}(s)} & \forall g\in G
\end{align}
which is to say that $g^{-1}(s)$ has the same orientation as $s$. 
On the other hand, since plaquettes $p$ reverse orientation under improper transformations, so too does $B_p$
\begin{align}
\sigma: B_p &\to B^\dagger_{|g^{-1}(p)|} & \forall \sigma \in G~\textrm{improper}
\end{align}
where  $|g^{-1}(p)|$ is the transformed plaquette but with a positive orientation. 
In short, this implies that improper transformation swap $B_p$ and $B^\dagger_p$. 

\paragraph{Reflection symmetry---} 
With these definitions, we find that $\HTC$ is symmetric under all proper transformations of the square lattice, but that the reflections $P$ transform the Hamiltonian by
\begin{align}
  P: \HTC(\Phi_e, \Phi_m) \rightarrow  \HTC(\Phi_e, - \Phi_m)\,.
\end{align}
There are two immediate consequences: (i) $\Phi_m \ne  0, \pi$ explicitly breaks reflection symmetry; (ii) the phase diagram of $\HTC$ has a symmetry $\Phi_m \to -\Phi_m$.

\paragraph{Charge conjugation---} 
We define charge conjugation $C$ by mapping the clock ($Z$) basis $\ket{s} \to \ket{-s~\textrm{mod}~p}$ on every edge. This induces
\begin{align}
C : \begin{cases}
    X_{e} \to X_e^{-1} \\
    Z_{e} \to Z_e^{-1} \\
    A_s \to A_s^{-1} \\
    B_p \to B_p^{-1}
\end{cases}
\end{align}
Thus, we obtain another unitary symmetry of the phase diagram:
\begin{align}
C : \HTC(\Phi_e, \Phi_m) \rightarrow \HTC(-\Phi_e, -\Phi_m) 
\end{align}
Or, combined with a reflection $\sigma$:
\begin{align}
C\sigma : \HTC(\Phi_e, \Phi_m) \rightarrow \HTC(-\Phi_e, \Phi_m) 
\end{align}

\paragraph{Time-reversal---}
We define time-reversal $\mathcal{T} = K$ to be complex conjugation with the clock basis $\ket{s}$ real. 
Accordingly,
\begin{align}
\mathcal{T} : \begin{cases}
            i \rightarrow -i \\
            X_e \rightarrow X_e \\
            Z_e \rightarrow Z^\dagger_e \\
            A_s \rightarrow A_s \\
            B_p \rightarrow B_p^\dagger 
        \end{cases}\,.
\end{align}
As a consequence, 
\begin{align}
\mathcal{T} : \HTC(\Phi_e, \Phi_m) \rightarrow \HTC(-\Phi_e, \Phi_m)\,.
\end{align}

Putting all of the discrete transformations together, we find that, no matter the couplings $\Phi_e, \Phi_m$, $\HTC$ has $CP\mathcal{T}$ symmetry.

\paragraph{Symmetry under global charge/flux shifts---}
The $\Zp$ model has several further useful unitary transformations associated with globally rotating the charge/flux ferris wheels by $2\pi/p$ on every site/plaquette.

Let us assume for simplicity that $p \vert L_x$ (similar constructions exist so long as $p \vert L_y$). 
Consider the unitary
\begin{align}
    U = \prod_{i_x=1}^{L_x} \prod_{i_y=1}^{L_y} Z_{\langle (i_x, i_y), (i_x+1, i_y)\rangle}^{i_x}\,.
\end{align}
This transformation clearly commutes with the plaquette and field terms.
However, it rotates the charge $A_s$ on every site $s$:
\begin{align}
    U: A_s \to \omega^{-1} A_s
\end{align}
Thus, 
\begin{align}
    U : \HTC(\Phi_e, \Phi_m) \to \HTC(\Phi_e + 2\pi/p, \Phi_m)
\end{align}
which induces a periodicity in the phase diagram of $\HTC$. 

A similar construction permits us to shift the flux $B_p$ globally. 
The corresponding unitary is
\begin{align}
    V = \prod_{i_x=1}^{L_x} \prod_{i_y=1}^{L_y} X_{\langle (i_x, i_y), (i_x,i_y+1) \rangle}^{i_x}\,
\end{align}
which transforms
\begin{align}
    V : B_p \to \omega^{-1} B_p\,.
\end{align}

Unlike the charge shifting operator $U$, the $V$ operator also modifies the field term proportional to $t_e$. 
Thus, the unperturbed phase diagram (\autoref{fig:phase_diagram_unpert} in the main text) is periodic in both $\Phi_e$ (by $U$) and $\Phi_m$ (by $V$) by $2\pi/p$, but when $t_e \ne 0$, the shift in $\Phi_m$ need not be a symmetry of the phase diagram.
Indeed, the Hall-on-Toric states do not appear in the absence of background flux (e.g., near $\Phi_m = 0$).

\section{Additional data}\label{sec:data}

\begin{figure}[t]
    \includegraphics[width=\columnwidth]{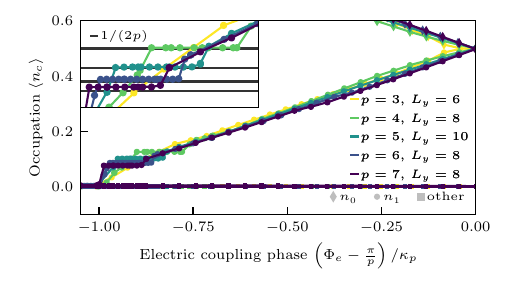}
    \caption{Charge occupation (\autoref{eq:nc} from the main text) computed with iDMRG (bond dimension $\chi=512$) for $t_e/\Gamma_e=0.05$ and different values of $p$ across the transition at $\Phi_e = \pi/p$ in units of $\kappa_p$ (\autoref{eq:kappa_p} from the main text). The markers represent the different charges.}
    \label{fig:charge_occupation_different_p}
\end{figure}

This appendix provides additional data on the charge occupation in the matter model (\autoref{eq:Hgauge} from the main text) for $(i)$ different values of $p$, $(ii)$ different ratios of $t_e/\Gamma_e$, and $(iii)$ varying circumferences $L_y$. Throughout, we fix the topological sector along the $y$-direction to $W_y = 1$. Calculations are performed in the Landau gauge oriented along the $x$-direction, such that the length ($L_x$) of the iDMRG unit cell is equal to $p$.

\paragraph*{(i)} We find that the Hall-on-Toric state remains stable for all values of $p > 3$. Its absence at $p=3$ ($p=2$ is the non-chiral Toric Code~\cite{kitaev_fault-tolerant_2003}) is consistent with results from the number-conserving case, where Hall states become unstable in the presence of strong flux~\cite{soerensen_fractional_2005,hafezi_characterization_2007,hafezi_fractional_2007,macaluso_charge_2020}. \autoref{fig:charge_occupation_different_p} displays the charge occupation for $p = 3, 4, 5, 6, 7$ at fixed $t_e/\Gamma_e = 0.05$. All cases (except for $p=3$) exhibit a plateau at the expected filling fraction $1/(2p)$ as illustrated in the inset.
\paragraph*{(ii)}

\begin{figure}[t]
    \includegraphics[width=\columnwidth]{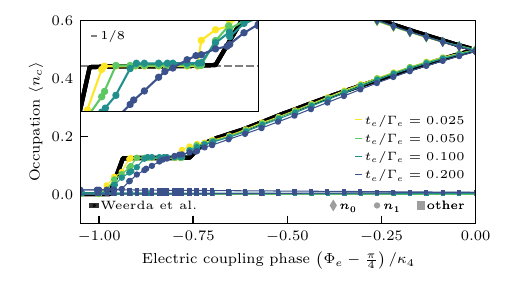}
    \caption{Charge occupation (\autoref{eq:nc} from the main text) computed with iDMRG (bond dimension $\chi=512$) for $p=4$ and different values of $t_e/\Gamma_e$ across the transition at $\Phi_e = \pi/4$ in units of $\kappa_p$ (\autoref{eq:kappa_p} from the main text). The markers represent the different charges. The black data is taken from Refs.~\cite{weerda_data_2023,weerda_fractional_2024} and computed for the number-conserving Bosonic Hofstadter problem.}
    \label{fig:charge_occupation_different_te}
\end{figure}

\begin{figure*}[t]
    \includegraphics[width=\columnwidth]{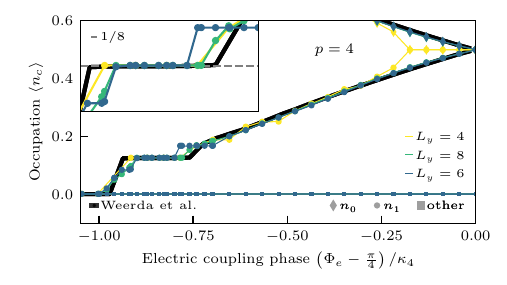}
    \includegraphics[width=\columnwidth]{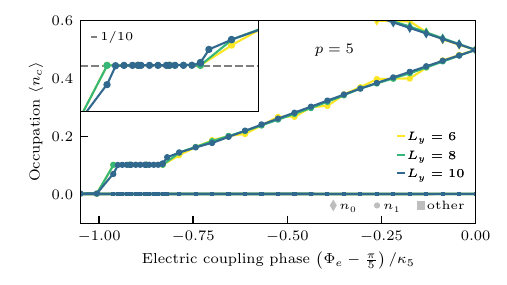}
    \caption{Charge occupation (\autoref{eq:nc} from the main text) computed with iDMRG (bond dimension $\chi=512$) for $p=4$ (left) and $p=5$ (right) and different values of $L_y$ across the transition at $\Phi_e = \pi/p$ in units of $\kappa_p$ (\autoref{eq:kappa_p} from the main text). We choose $t_e/\Gamma_e=0.05$. The markers represent the different charges. The black data for $p=4$ is taken from Refs.~\cite{weerda_data_2023,weerda_fractional_2024} and computed for the number-conserving Bosonic Hofstadter problem.}
    \label{fig:charge_occupation_finite_size}
\end{figure*}

The perturbative interpretation of number-conserving charges breaks down at larger values of $t_e/\Gamma_e$, driving the system out of the gapped phases (both $\mathbb{Z}_p$ and Hall-on-Toric). \autoref{fig:charge_occupation_different_te} shows the charge occupation for varying values of $t_e/\Gamma_e$ at $p=4$. In the limit $t_e/\Gamma_e \rightarrow 0$, we expect the model to recover the Bosonic Hofstadter model (data from Refs.~\cite{weerda_data_2023,weerda_fractional_2024} is shown in black).

As $t_e/\Gamma_e$ increases, the width of the charge plateau shrinks, and for $t_e/\Gamma_e \approx 0.2$, the plateau vanishes. At these values, the occupations of other charge sectors ($n_c$ for $c\neq 0,1$) begin to increase, suggesting the emergence of a more complex, competing state. This motivates the fading color scheme used in \figref{fig:1}{c}, which signals the breakdown of the Hall-on-Toric state.

\paragraph*{(iii)} The Hall-on-Toric states remain stable across different circumferences, as demonstrated in \autoref{fig:charge_occupation_finite_size} for $p = 4$ and $p = 5$. Although a precise scaling analysis of the plateau width with system size is not feasible, we are confident that the plateau persists in the thermodynamic limit. 

Notably, when the circumference matches the magnetic unit cell length ($L_y = p$), additional plateaus appear at filling fractions consistent with Bosonic Hall states (e.g., $\nu=2$ with $n_1=2/5$ in \figref{fig:charge_occupation_finite_size}{b}). In contrast to the $\nu = 1/2$ phase, these higher filling states become unstable as the circumference changes. We believe that this is due to the formation of charge density waves, as outlined in Refs.~\cite{tao_fractional_1983,rezayi_laughlin_1994,bergholtz_pfaffian_2006,seidel_abelian_2006,palm_bosonic_2021,weerda_fractional_2024}.

\section{$\Zp$ Hall conductance}\label{app:Hall}

We generalize the charge pumping~\cite{laughlin_quantized_1981} for fractional quantum Hall states to the non-number-conserving matter model. The ground state is represented as an infinite matrix product state, and the Schmidt values at a bipartition can be organized according to the underlying symmetries of the system. In the $\Zp$ case, the defining symmetry corresponds to particle number conservation modulo $p$, while it is the $U(1)$ symmetry in the conventional setting of fractional quantum Hall states.
\begin{align}
    \ket{\Psi} := \sum_c\sum_{i=1}^{\chi_c}\Lambda_{(c,i)} \ket{\psi_{(c,i)}^L} \otimes \ket{\psi_{(c,i)}^R}\,.
\end{align}
We have dropped the summation over $k_y$. Here, $c$ refers to the charge difference between the two sides of the cylinder. $c$ can take any integer in the $U(1)$ case and is restricted to integers modulo $p$ for the $\Zp$ case: $\{0,...,p-1\}$. Note that additional charges can be placed on one side of the infinite cylinder such that the weights can be shifted.

We define the weight associated with each charge channel by:
\begin{align}
    W_c := \sum_{i=1}^{\chi_c}\Lambda_{(c,i)}^2\,.
\end{align}
The sum over all the weights is one due to the normalization of the wavefunction: 
\begin{align}
    \sum_cW_c=\braket{\Psi}{\Psi}=1\,.\label{eq:norm}
\end{align}

Starting from the ground state of the respective models, we adiabatically thread $\Theta_y$ through the periodic (finite) direction. The flux is inserted via the horizontal edges of the lattice. 
The Hamiltonian as a function of $\Theta_y$ is
\begin{align}
    \HCM(\Theta_y) =& -\Gamma_e\sum_{ (x,y)} e^{-i\Phi_e}{C}_s+\text{h.c.}\label{eq:App_Hgauge_Theta}\\
    &- t_e\sum_{ (x,y)} \omega^x\; {R}^\dagger_{(x,y)}{R}_{(x,y+1)}+\text{h.c.}\nonumber\\
    &- t_e\sum_{ (x,y)} e^{i\Theta_y/L_y}\; {R}^\dagger_{(x,y)}{R}_{(x+1,y)}+\text{h.c.}\,,\nonumber
\end{align}
where the sums run over all lattice sites denoted by their $x$ and $y$ coordinates: $(x,y)$.
This agrees with the definition of $\tau_{ij}$ from \autoref{eq:tau_def} for $\Theta_y=0$.

\begin{figure}[t]
    \centering
    \includegraphics[width=1\columnwidth]{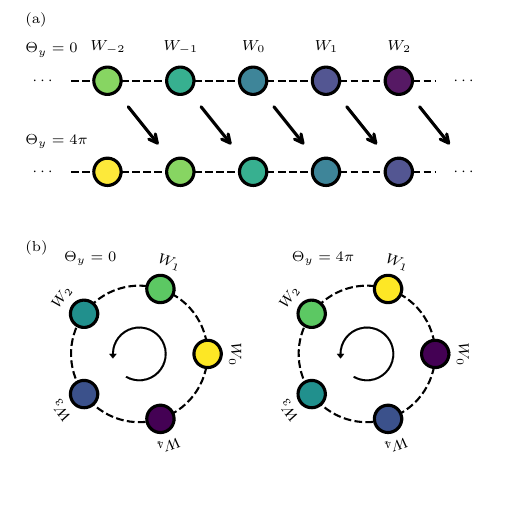}
    \caption{Illustration of the weight shift in \autoref{eq:charge_shift} for the number-conserving model in (a) and in \autoref{eq:charge_shift_mod_p} for the matter model in (b). For the number-conserving model, the shift is uniform, and weights $W_c$ can take any integer value. Adiabatically threading flux of $4\pi$ can be understood as a one-dimensional, discrete transport process: $W_c(0)\rightarrow W_{c+1}(4\pi)$. The transport in the $\Zp$ model results into a cyclic shift: $W_c(0)\rightarrow W_{\modp{c+1}}(4\pi)$.}
    \label{fig:Hall_cartoon}
\end{figure}
\begin{figure}[t]
    \includegraphics[width=\columnwidth]{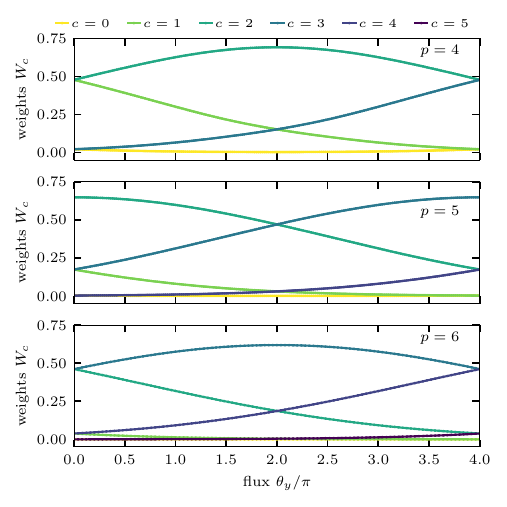}
    \caption{Weights during the adiabatic evolution for different charge channels in the $\mathbb{Z}_p$ model. The iDMRG unit cell has the size $L_x=p$ and $L_y=2p$. The adiabatic evolution from $\Theta_y = 0$ to $\Theta_y = 4\pi$ was performed using 201 steps for $p = 4$ and $p = 5$, and 101 steps for $p = 6$ with a bond dimension $\chi = 1024$.}
    \label{fig:adiabatic_evolution}
\end{figure}

We obtain the ground state at $\Theta_y=0$ by fixing the particle number to match $\nu=1/2$ filling for the hardcore Boson model. In the $\Zp$ case, we select parameters that place the system within the Hall-on-Toric phase. The ground state is adiabatically evolved by varying $\Theta_y$.

In the number-conserving model, weights associated with each charge sector undergo a uniform shift during the adiabatic evolution:
\begin{align}
    W_c(\Theta_y = 0) = W_{c+1}(\Theta_y = 4\pi)\,.\label{eq:charge_shift}
\end{align}
Hence, the pumping can be understood as a one-dimensional, discrete transport process. It is illustrated in \figref{fig:Hall_cartoon}{a}, which displays a subset of weights and their evolution after threading $\Theta_y = 4\pi$ through the system. For example, the green weight initially associated with $W_{-2}$ at $\Theta_y = 0$ shifts to $W_{-1}$ at $\Theta_y = 4\pi$.

In the $\Zp$ case, we find that weights are shifted modulo $p$:
\begin{align}
    W_c(\Theta_y=0)= W_{\modp{c+1}}(\Theta_y=4\pi)\,.\label{eq:charge_shift_mod_p}
\end{align}
This behavior is illustrated in \figref{fig:Hall_cartoon}{b} for $p = 5$. The weight associated with $W_0$ (yellow) shifts to $W_1$ after threading $\Theta_y = 4\pi$. However, in contrast to the $U(1)$ case, the weight of $W_4$ (dark blue) wraps around and shifts to $W_0$.
The adiabatic evolution is depicted \autoref{fig:adiabatic_evolution}.

We further show numerical values of the weights for both models in \autoref{tab:wide_table} at $\Theta_y=0$ and $\Theta_y=4\pi$. To probe the uniform shift in the $U(1)$ case, \autoref{eq:charge_shift}, and a cyclic shift in the $\mathbb{Z}_p$ case, \autoref{eq:charge_shift_mod_p}, we compute the following quantities:
\begin{align}
    \Delta_{\Zp}&:= \sqrt{\sum_c \left(W_c(0)- W_{\modp{c+1}}(4\pi)\right)^2}\,.\\
    \Delta_{U(1)}&:= \sqrt{\sum_c \left(W_c(0)- W_{c+1}(4\pi)\right)^2}\,.
\end{align}

In both models, we observe excellent agreement: $\Delta_{U(1)}$ and $\Delta_{\mathbb{Z}_p}$ vanish within numerical accuracy for $p = 4, 5, 6$. In the $U(1)$ model, we observed that the weights are concentrated within $p$ consecutive charge sectors (values below $10^{-4}$ have been discarded in \autoref{tab:wide_table}). Moreover, the numerical values of the weights for both models closely match within this threshold, suggesting that they share the same ground-state structure.

\begin{table*}[b]
    \centering
    \begin{tabular}{ c | c | c || c | c | c | c | c | c | c || c }
        $p$ & Model & $\Theta_y$ &  $W_0$ & $W_1$ & $W_2$ & $W_3$ & $W_4$ & $W_5$ & $W_6$ & $\Delta_{U(1)/\Zp}$\\ \hline
        \multirow{4}{*}{4} & \multirow{2}{*}{$U(1)$} & $0$ & 0.0220 & 0.4779 & 0.4779 & 0.0220 & -- & -- & -- & \multirow{2}{*}{$2\cdot 10^{-6}$} \\
        &  & $4\pi$  &  -- & 0.0220 & 0.4779 & 0.4779 & 0.0220 & -- & --  & \\\cline{2-11}
        & \multirow{2}{*}{$\Zp$} &$0$ & 0.0222 & 0.4778 & 0.4778 & 0.0222 & --  & --  & --   & \multirow{2}{*}{$68\cdot 10^{-6}$}\\
        & & $4\pi$ & 0.0222 & 0.0222 & 0.4778 & 0.4777 & --  & --  & --  &  \\\hline
        \multirow{4}{*}{5} &  \multirow{2}{*}{$U(1)$} & $0$ & 0.0022 & 0.1736 & 0.6484 & 0.1736 & 0.0022 & -- & --  & \multirow{2}{*}{$2\cdot 10^{-6}$} \\
        &  & $4\pi$ &  -- & 0.0022 & 0.1736 & 0.6484 & 0.1736 & 0.0022 & --  &  \\\cline{2-11}
        & \multirow{2}{*}{$\Zp$} &0 & 0.0026 & 0.1736 & 0.6476 & 0.1736 & 0.0026 & --  & --  & \multirow{2}{*}{$83\cdot 10^{-6}$}\\
        &  & $4\pi$ & 0.0026 & 0.0026 & 0.1736 & 0.6477 & 0.1736 & --  & --  &  \\\hline
        \multirow{4}{*}{6} & \multirow{2}{*}{$U(1)$} &$0$ &  0.0001 & 0.0385 & 0.4614 & 0.4614 & 0.0385 & 0.0001 & --  & \multirow{2}{*}{$4\cdot 10^{-6}$}\\
        & & $4\pi$ & --  &  0.0001 & 0.0385 & 0.4614 & 0.4614 & 0.0385 & 0.0001  & \\\cline{2-11}
        & \multirow{2}{*}{$\Zp$}&$0$  & 0.0003 & 0.0385 & 0.4612 & 0.4612 & 0.0384 & 0.0003 & -- & \multirow{2}{*}{$573\cdot 10^{-6}$} \\
        & & $4\pi$ &  0.0003 & 0.0003 & 0.0386 & 0.4616 & 0.4609 & 0.0382 & --  &  \\
    \end{tabular}
    \caption{Weights for different charge channels in the $U(1)$ and $\mathbb{Z}_p$ cases are shown. The iDMRG unit cell has the size $L_x=p$ and $L_y=2p$. The adiabatic evolution from $\Theta_y = 0$ to $\Theta_y = 4\pi$ was performed using 201 steps for $p = 4$ and $p = 5$, and 101 steps for $p = 6$. Throughout, the bond dimension was fixed at $\chi = 1024$. In the $U(1)$ case, we have discarded weights smaller than $10^{-4}$. Since the system is placed on an infinite cylinder, additional charges can be placed on one side of the cylinder such that the weights can be shifted for the initial wavefunction at $\Theta_y=0$.}
    \label{tab:wide_table}
\end{table*}

\end{document}